\documentclass{elsart}
\usepackage{graphicx,amssymb,lineno, aas_macros}

\makeatletter
\def\elsartstyle{%
    \def\normalsize{\@setfontsize\normalsize\@xiipt{14.5}}
    \def\small{\@setfontsize\small\@xipt{13.6}}
    \let\footnotesize=\small
    \def\large{\@setfontsize\large\@xivpt{18}}
    \def\Large{\@setfontsize\Large\@xviipt{22}}
    \skip\@mpfootins = 18\p@ \@plus 2\p@
    \normalsize
}
\@ifundefined{square}{}{\let\Box\square}
\makeatother

\newcommand{\eg}{\emph{e.g.}}

\begin{document}

\begin{frontmatter}
\title{Measurements of cosmic-ray secondary nuclei at high energies with the first flight of the CREAM balloon-borne experiment}

\author[umd]{H.S. Ahn},
\author[osu]{P.S. Allison},
\author[uos]{M.G. Bagliesi},
\author[osu]{J.J. Beatty},
\author[uos]{G.~Bigongiari},
\author[uoc]{P.J. Boyle},
\author[osu]{T.J. Brandt},
\author[uom]{J.T. Childers},
\author[psu]{N.B. Conklin},
\author[psu]{S. Coutu},
\author[uom]{M.A. Duvernois\thanksref{haw}},
\author[umd]{O.~Ganel},
\author[ewu]{J.H. Han},
\author[ewu]{H. J. Hyun},
\author[ewu]{J.A. Jeon},
\author[umd]{K.C. Kim},
\author[ewu]{J.K. Lee},
\author[umd]{M.H. Lee},
\author[umd]{L. Lutz},
\author[uos]{P.~Maestro},
\author[umd]{A. Malinin},
\author[uos]{P.S. Marrocchesi},
\author[ksu]{S.A. Minnick},
\author[psu]{S.I. Mognet},
\author[ewu]{S. Nam},
\author[nku]{S.L. Nutter},
\author[ewu]{I.H. Park},
\author[ewu]{N.H. Park},
\author[umd]{E.S. Seo},
\author[umd]{R. Sina},
\author[uoc]{S.P.~Swordy\corauthref{cor}},
\author[uoc]{S.P.~Wakely},
\author[umd]{J. Wu},
\author[ewu]{J. Yang},
\author[umd]{Y.S. Yoon},
\author[uos]{R. Zei},
\author[umd]{S.Y. Zinn}

\address[umd]{Institute for Physical Science and Tech., University of Maryland, College Park, MD 20742, USA}
\address[osu]{Department of Physics, Ohio State University, Columbus, OH 43210, USA}
\address[uos]{Department of Physics, University of Siena \& INFN, Via Roma 56, 53100 Siena, Italy}
\address[uoc]{Enrico Fermi Institute and Dept. of Physics, University of Chicago, Chicago, IL 60637, USA}
\address[uom]{School of Physics and Astronomy, University of Minnesota, Minneapolis, MN 55455, USA}
\address[psu]{Pennsylvania State University, University Park, PA 16802, USA}
\address[ewu]{Department of Physics, Ewha Womans University, Seoul 120-750, Republic of Korea}
\address[ksu]{Department of Physics, Kent State University, Tuscarawas, New Philadelphia, OH 44663, USA}
\address[nku]{Department of Physics, Northern Kentucky University, Highland Height, KY 41099, USA}

\thanks[haw]{now at the University of Hawaii}
\corauth[cor]{Corresponding author. Address: s-swordy@uchicago.edu}

\begin{abstract}

We present new measurements of heavy cosmic-ray nuclei at high energies performed during the first flight of the balloon-borne
cosmic-ray experiment CREAM (Cosmic-Ray Energetics And Mass).  This instrument uses multiple charge detectors and a transition
radiation detector to provide the first high accuracy measurements of the relative abundances of elements from boron to oxygen up to
energies around 1 TeV/n. The data agree with previous measurements at lower energies and show a relatively steep decline ($\sim
E^{-0.6}$ to $E^{-0.5}$) at high energies. They further show the source abundance of nitrogen relative to oxygen is $\sim 10\%$ in the
TeV/n region.

\end{abstract}

\end{frontmatter}

\section{Introduction}
\label{intro}

The first flight of the Cosmic-Ray Energetics and Mass (CREAM) high-altitude balloon experiment took place in Antarctica during
December 2004 and January 2005. A central goal of this flight was to investigate the source properties of cosmic rays through the
careful measurement of those cosmic nuclei which are predominantly of secondary origin.

The escape of particles from the Galaxy during propagation is known to be energy dependent at high energy.  As a result, the spectrum
of cosmic rays measured at Earth is different from the spectrum of these particles at their source.  Understanding the magnitude of
this difference is crucial for solving the puzzle of cosmic-ray origins. Secondary nuclei are particularly useful in addressing this
goal because they are produced largely by the spallation interactions of primary particles as they propagate from their source regions
through the interstellar medium to Earth. The pattern of relative abundances of secondary and primary nuclei therefore encodes
information about the propagation history of the primary cosmic rays, allowing us to identify the contribution from propagation to the
spectrum observed at earth.  A recent review has identified the measurement of secondary nuclei abundances as a key goal for achieving
future progress in cosmic-ray science \cite{2007ARNPS..57..285S}.

It is known from earlier measurements (see, \eg,
\cite{1972PhRvL..29..445J,1973ApJ...180..987S,1990ApJ...349..625S,1990A&A...233...96E}) that the diffusion escape time for particles
from our Galaxy decreases with increasing particle energy, or magnetic rigidity. This decrease can be described as a power law above
some threshold rigidity near 2-10~GV.  Below this rigidity, the escape time is roughly constant and the interstellar pathlength is
approximately proportional to the particle velocity $\beta$(=v/c)\cite{2000AIPC..528..421D}. Above 10~GV, the escape time can be
characterized as a pathlength (in g/cm$^2$) by assuming a constant density for the interstellar medium.  A typical form for the
rigidity dependence of this quantity is $\lambda = {\lambda}_0(R/R_0)^{-\delta}$.  Here $\lambda$ is the mean escape pathlength, $R$
is the nucleus magnetic rigidity, and $\lambda_0$($\sim$10g/cm$^2$) is the pathlength at the threshold rigidity $R_0$ (typically
10~GV).

The variable $\delta$ is the key parameter in the description of cosmic-ray diffusion from the Galaxy at high energies. In the
simplest models of cosmic-ray transport, primary particles from a source with spectral index $\alpha$ escape more easily at high
energy to produce an steepening in the slope of the energy spectrum observed at Earth to an index of $\alpha + \delta$.

The propagation of cosmic rays in our Galaxy is an essentially statistical process at energies where the radius of gyration in typical
bulk fields is significantly less than the size of randomness in the magnetic field structures. In practice, this means that at the
energies where the measurements of secondary nuclei have thus far been possible ($<$100~GeV/n) the propagation is indeed
well-approximated by models where cosmic rays simply diffuse out of the Galaxy.  As a result, the pathlength distribution for each
individual nuclear species is close to an exponential in form, with a mean value at high energy given by the expression above.

The standard procedure when measuring secondary nuclei is to compare their abundances to those of their most likely (primary) parent
nuclei, and examine how these ratios vary with energy.  This can determine the value of the escape energy dependence, $\delta$. For
example, the ratio of boron (essentially all of which is secondary) to its most likely parent nucleus, carbon, is expected to decline
with energy with a slope which almost directly yields a value for $\delta$ above $\sim$~30 GeV/n. nitrogen, on the other hand, is only
partly secondary; it is known to have a residual abundance at the source. The ratio of nitrogen to oxygen is expected to decline less
rapidly with energy. Finally, the ratio of the two primary nuclei carbon and oxygen is expected to remain relatively constant with
energy.

In this work, we describe measurements of the ratios of boron to carbon, nitrogen to oxygen, and carbon to oxygen over an energy range
of ~1GeV/n-1TeV/n. Heavier nuclei (Z$>$8) will be the subject of future publications.  One crucial aspect of these measurements is the
correction for secondary particle production in the local atmospheric overburden above the balloon instrument. The systematic error
associated with this correction becomes significant at high energies and ultimately imposes a limit for the maximum energy at which
such measurements might be pursued on a balloon flight.

\section{CREAM Instrument and Flight}\label{instrument}

The CREAM instrument was first launched from Williams Field near McMurdo Station in Antarctica on December 16th, 2004. The payload
circled the South Pole three times in a flight which lasted nearly 42 days\cite{2005ICRC-CREAM}.  The mean altitude during the flight
was 38.6km, corresponding to an average atmospheric overburden of 3.9 g/cm$^2$.  The data for the present study were provided by a
trigger scheme designed to tag heavy nucleus events (charge Z$\gtrsim$3). We term these triggers ``Hi-Z" events, around 15 million of
these events are used in the present analysis.

\begin{figure}
  \centering
  \includegraphics[width=0.55\linewidth]{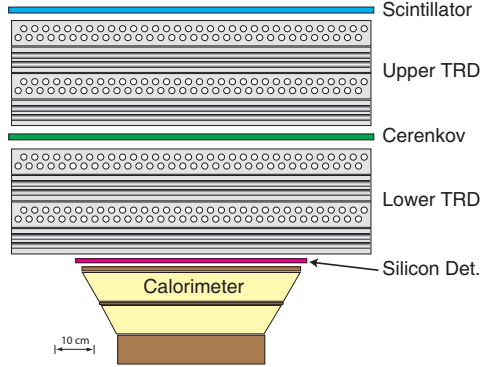}
  \caption{A Schematic view of the CREAM instrument for the first flight.}
  \label{fig:Schematic}
\end{figure}

The instrument configuration used for this flight is shown schematically in Figure \ref{fig:Schematic}. A much more detailed
description of the instrument, along with a discussion of calibration processes, detector performance, and general architecture is
given in a recent instrument paper \cite{2007NIMPA.579.1034A}.  Here we briefly review only those detector elements used for the Hi-Z
analysis.

The detector system consists of a stack of horizontal components resting on a solid base plate. Importantly, there is no pressure
gondola shell for CREAM.  Not only does this significantly decrease the mass of the payload, it reduces the material overburden of the
instrument, limiting the probability of local interactions of cosmic-ray nuclei.  Such interactions can cause backgrounds of secondary
particles via charge-changing nuclear processes.

At the base of the detector stack is a hadronic calorimeter (CAL) which has been described in detail elsewhere
\cite{2006NuPhS.150..272A}.  The calorimeter is designed to measure particles at the highest energies and is not used in the present
analysis.  Above the CAL is a silicon charge detector (SCD) consisting of 2912 individual pixels, each with an area of $1.55 \times
1.37$ cm$^2$\cite{2007NIMPA.570..286P}.  The pixels overlap slightly to provide full coverage over a total area of $78 \times 79$
cm$^2$. The dynamic range of the readout system can accommodate nuclei from $Z=1$ to $Z=28$.  Over this range the detectors have a
remarkably linear response.

Above the SCD lies a precision gas transition-radiation detector (TRD)\cite{2002APh....18...67W}, split into two segments.  Each
segment consists of blocks of plastic foam ($120 \times 120$ cm$^2$ in area) in which are embedded 2~cm diameter thin-walled
proportional tubes.  There are 256 such tubes in a segment, each filled with a 95\% xenon 5\% methane gas mixture.  The CREAM
coordinate system is defined with the $z$ direction vertical in Figure \ref{fig:Schematic}, with perpendicular $x$ and $y$ directions
corresponding to the major axes of the detector planes. The TRD tubes in each segment are divided into 4 sets of 64, each of which is
alternatively aligned with either the $x$ or $y$ axis.  This provides a view of the particle track in both the $xz$ plane and the $yz$
plane.

Each proportional tube in the array is instrumented with a dual-gain amplifier system which provides a dynamic range of
$\sim$4000. This allows for the measurement of nuclei from $Z=3$ to $Z=26$.  Between the two TRD segments is an acrylic
Cherenkov detector (CD), which is instrumented by eight photomultiplier tubes (PMTs) viewing wave-shifter bars along
the edges of the plastic. The central radiator has a refractive index near 1.5 and is doped with a wave-shifting dye to
provide near-isotropic emission within the volume.

At the top of the instrument is a plastic scintillation assembly, the Timing Charge Detector (TCD). The detector consists of eight
$120\times 30 \times 0.5$cm$^3$ paddles, viewed at each end by fast photomultipliers. They are arranged in two orthogonal layers of
four paddles each, providing a coverage area of $120 \times 120$ cm$^2$.  Both the timing and amplitude of the scintillator signals
from each event are digitized by these detectors. In this work, however, we only use the event amplitudes. For these, the TCD has a
dynamic range for nuclei from $Z=1$ to $Z=28$.  The total grammage of the detector stack, from TCD to SCD is about 7.0 g/cm$^2$.

The nuclei presented in this paper are almost all in the energy range below 1~TeV/n. The energy measurement device used in this regime
is the TRD.  Although the Calorimeter trigger threshold accepts events near the upper end of this energy regime, it is not included in
the present analysis since it restricts the available collection aperture. While the amount of transition radiation produced at this
energy is not large (the nominal threshold for production is around Lorentz factors $\gamma > 10^3$), the proportional tube gas,
xenon, has one of the largest known relativistic rises in ionization energy loss\cite{WalentaRelRise79}.  This allows for an estimate
of particle $\gamma$ in the range $4 < \gamma < 5\times 10^2$. Importantly, the response of this detector has been calibrated during
beam tests at CERN. This is discussed extensively elsewhere \cite{2003ICRC....4.2233S}.

\section{Data Overview}
In this section we describe in some detail the data collected during the flight, as well as some of the corrections applied to those
data. A Hi-Z event is required to generate a signal above a fixed threshold in both the CD and TCD within a coincidence window of
$\sim 100$ns. These thresholds were adjusted before the flight to ensure high efficiency for relativistic boron ($Z=5$) nuclei, while
rejecting the tail of the much more abundant Helium nuclei ($Z=2$).

\begin{figure}[tb]
  \centering
  \includegraphics[width=0.99\linewidth]{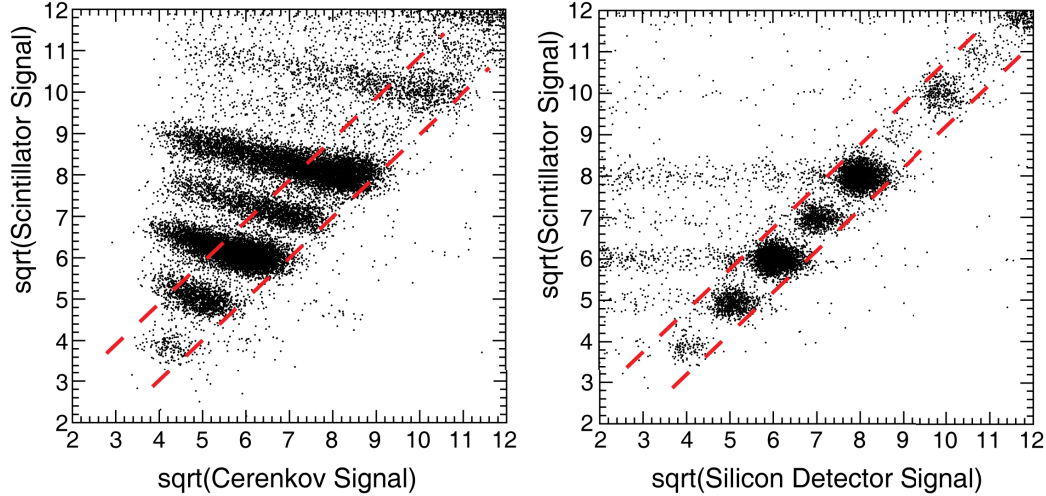}
\caption{Scatter plots of detector signals for a small fraction of the flight
data. Left: The square root of the TCD signal versus the square root of the CD
signal.  Also shown are representative cuts applied to select relativistic
nuclei (between dashed lines).  The vertical scale here has been adjusted to
approximate the nuclear charge and all signals are corrected for gain and
pathlength.  Right: The square root of the corrected TCD signal versus the
square root of the corrected SCD for the selected relativistic nuclei from the
left panel.  Again, representative cuts are shown which select nuclei that do
not undergo charge-changing interactions while traversing the
instrument.}\label{fig:Scatters}
\end{figure}

\subsection{Detector Corrections}
One of the benefits of a tracking instrument is that it provides a simple and direct mechanism for identifying and
correcting non-uniformities in detector response.  After constructing particle trajectories using the tracking
algorithm described below, we apply mapping corrections for every event collected.  The maximum level of these
corrections is $\sim$50\% for the TCD and $\sim$30\% for the CD. No positional mapping is required for the silicon
detector, but the individual pixels do receive a gain-matching correction.  Again using tracking information, the
signals from all the detector systems are also corrected for pathlength.

In the TCD, an additional correction is made for the (Birk's Law) saturation in the scintillator and non-linearity of the PMT
response.  The TRD's proportional tube signals are also corrected for gain variations caused by fluctuations in pressure and
temperature.  These signals are gain-matched to a precision of $\sim$3\% over the entire flight. Additionally, there were several
calibration procedures implemented periodically throughout the flight, as discussed in \cite{2007NIMPA.579.1034A}.

\subsection{Charge Identification}

To outline the analysis procedures, we first discuss in general terms the identification of charges using the CREAM detectors.  This
initial estimate of charge is used as an input to one of the tracking algorithms.   Later, in Section \ref{sec:Anal}, we describe the
full relativistic energy analysis which requires more detailed interpretation of the events. The left-hand panel of Figure
\ref{fig:Scatters} shows a small sample of events measured in the TCD and CD after applying the above corrections. The individual
charges are clearly visible. The vertical TCD scale is arbitrary, but has been adjusted to approximate the derived charge of the
incoming relativistic nuclei. Also apparent on this plot is the threshold setting for the CD (abscissa), which can be seen to admit
relativistic boron nuclei, and with lesser efficiency, Beryllium.

In a polar flight, the geomagnetic cutoff rigidity is very low, admitting a large flux of low-energy particles.  Since we are
primarily interested in high-energy particles for this study, we use the Cherenkov Detector to identify and reject the low energy
ones. The CD has a refractive index of 1.5, which provides a threshold velocity of $\beta\sim 0.7$.  Particles with low CD signal are
therefore easily identified as having an energy of a few hundred MeV/n, well below minimum ionization in the proportional tubes of the
TRD.  An example of the cuts used to reject such nuclei are shown in the left panel of Figure \ref{fig:Scatters} (dashed lines).

Also shown in Figure \ref{fig:Scatters} (right panel) is a scatter plot between the TCD and the SCD signals (after the above cuts are
applied).  In general, the charge measurements by these devices are in good agreement.  However, there are a number of events in which
the charge measured by the SCD (abscissa) is lower than that measured by the TCD.  These are caused by nuclear interactions in the
instrument. Such events represent a small fraction of the data\cite{2007ICRC-TCD} and can be easily removed from the data by a simple
consistency check between the TCD and SCD signal. A representative cut is shown by the dashed lines on the figure.

\begin{figure}[tb]
  \centering
  \includegraphics[width=0.99\linewidth]{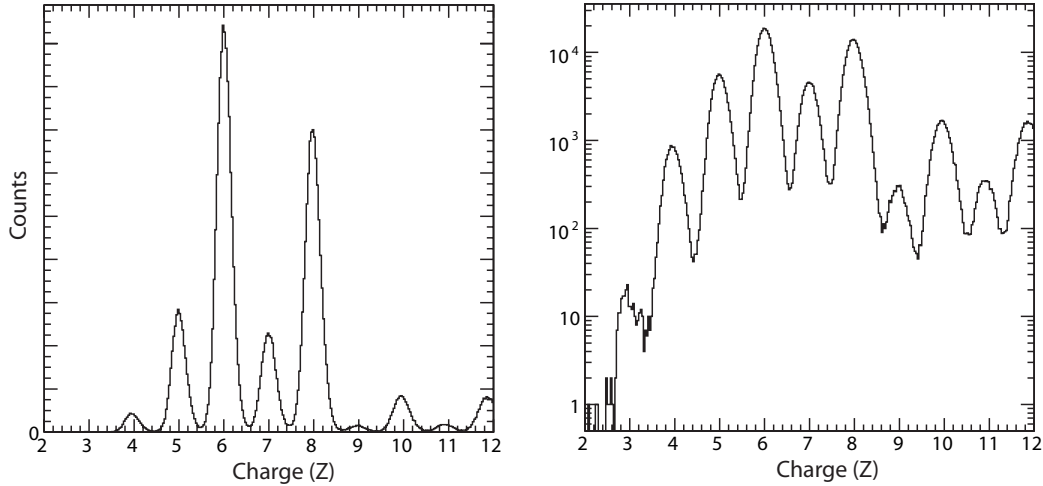}
  \caption{Charge histogram of nuclei with Hi-Z triggers.  The charge estimate shown is the mean of the square
  root of the corrected TCD signal and the square root of the corrected SCD signal.  The data were selected using the
  cuts from Figure \ref{fig:Scatters}.  The left-hand panel has a linear scale and the right-hand shows the same data with
  a logarithmic scale.  The distribution for carbon has a width of $\sigma_Z$=0.16 charge units.}\label{fig:Charge}
\end{figure}
After the relativistic particle cuts are applied to the CD data and the consistency cut has been applied between the TCD and SCD
signals, the overall charge resolution can be determined.  Figure \ref{fig:Charge} shows the mean of the SCD and TCD signals
($(\sqrt{TCD}+\sqrt{SCD})/2$) for relativistic Hi-Z, non-interacting particles collected during the flight.  The charge distribution
extends down to Lithium, which produces Hi-Z triggers at low efficiency because of the fixed CD threshold discussed above.  Across the
whole charge spectrum, all of the nuclei are visibly separated from their neighbors, including the rarer nuclei, Fluorine (Z=9) and
Sodium (Z=11).  Overall, the achieved charge resolution is excellent, with an RMS ($1\sigma$) resolution for carbon nuclei of
$\Delta$Z = 0.16 charge units. Performance at this level is crucial for the successful study of rare secondary particles, as discussed
below.

\section{Analysis Methods\label{sec:Anal}}

In the following section we will discuss the analysis procedures and corrections used to reduce the data from the CREAM
flight.  This includes a discussion of the final selection cuts (which are somewhat more restrictive than the cuts
discussed above), the energy-determination mechanism, and energy deconvolution procedures.  We begin with a synopsis of
the track-fitting algorithms.

\subsection{Event Track Fitting}

The heart of the analysis is the track fit.  The results of this fit are used in the mapping corrections described above to correct
detector signals.  They are also used to determine the actual track length of the particle in each of the TRD tubes it traverses.  The
accuracy of this information is crucial for properly determining the $dE/dx$ signal for each event. Because this number needs to be
determined as well as possible, the fitted tracks must have a precision significantly less than a tube diameter.  This is achieved by
a multistage track-fitting procedure.

The first-stage track fitter employs a simple and very fast least-squares minimization which ignores the physical extent of the
proportional tubes.  Outlying signals are rejected from the fit by iteratively contracting an envelope around the estimated track
position. Within 3 passes, the precision of this initial fit is $\sim$5mm FWHM in both the $xy$ and $yz$ planes.  The fit is then
improved using a more sophisticated second-level algorithm which accounts for the full three-dimensional structure of the TRD.  This
algorithm uses the Minuit software package to perform a maximum-likelihood fit incorporating the different energy-loss distributions
produced by tracks at different impact parameters. Because these distributions also vary with particle charge, an initial
determination of particle charge Z (discussed above) is used as input to the algorithm.  When initialized with the parameters provided
by the level-one track, this fitter achieves an accuracy at the level of $\sim$1 mm RMS.

\begin{figure}
  \centering
  \includegraphics[width=0.99\linewidth]{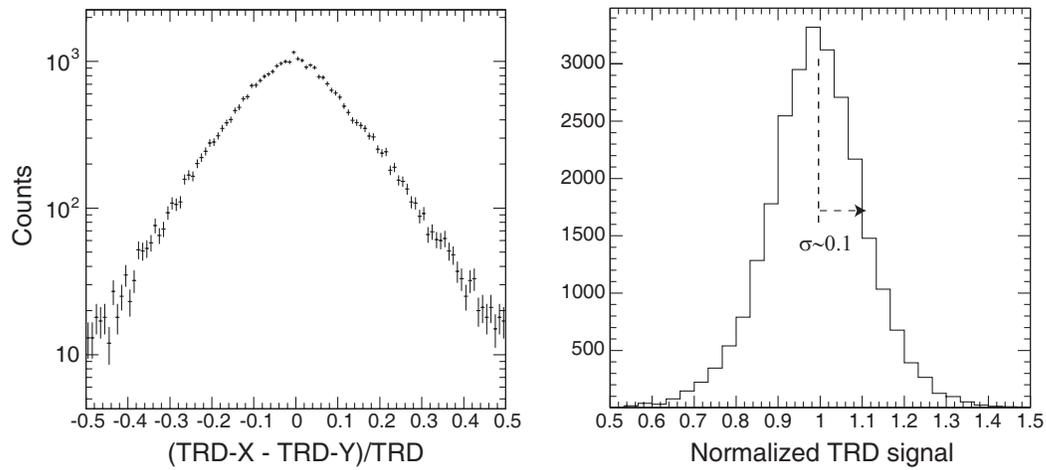}
\caption{Left: ~The fractional difference between the TRD signal determined from the $xz$ track and the $yz$ track for oxygen nuclei.
The result has an RMS value of $\sim$14\%. Right:~The TRD sum signal distribution for flight oxygen nuclei, selected to be at minimum
ionization using the CD. The RMS resolution is $\sigma\sim $10\%.}\label{fig:sigdist}
\end{figure}

Using the tracking information, the TRD's $dE/dx$ signal is directly derived from the individual tube signals.  A quality cut is
applied which requires individual track segments to be at least 1.5~cm long, helping to minimize the impact of fluctuations for thin
gas layers.  A simple test of the accuracy of this process can be obtained by comparing the signal derived from the $xz$ plane to that
of the $yz$ plane. These fits are essentially independent except for the common track energy-loss parameter, and therefore a
comparison gives a rough estimate of the overall error in the determination of the $dE/dx$ value.  The left panel of Figure
\ref{fig:sigdist} shows such a test for flight oxygen nuclei. The relative RMS difference between the $xz$ and $yz$ determinations is
$\sim 14\%$, implying that the resolution in the \emph{combined} TRD is $\sigma\sim$7\%.  The actual resolution, discussed below, is a
little broader than this because of the correlations introduced by the track energy-loss parameter.

\subsection{Energy Determination}

The energy of incoming particles can be determined in two ways.  At lower energies, the signal from the Cherenkov Detector can be used
to estimate velocity, and hence energy.  The TRD proportional tubes are used at energies above where the CD response has saturated.
The signal in these tubes is produced by a combination of two energy-loss mechanisms, both of which scale with the energy of the
nucleus. The primary component comes from Coulomb collisions of the nuclei with electrons in the gas.  An additional contribution
comes from any x-ray transition radiation absorbed in the xenon gas mixture.  In the energy region of the present measurements
($\lesssim 1$ TeV/n), the tube signals are essentially all due to ionization energy loss.

\begin{figure}
  \centering
  \includegraphics[width=0.90\linewidth]{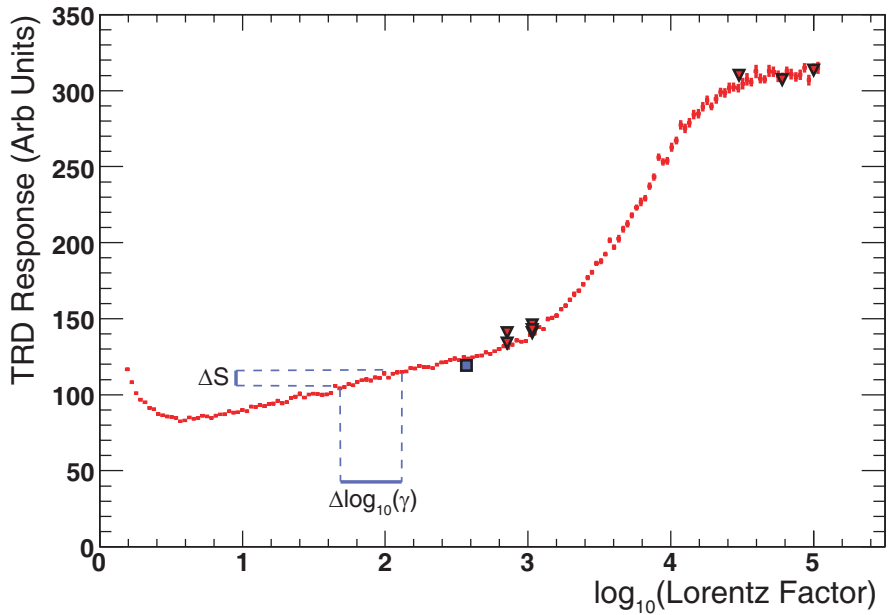}
\caption{The response of the TRD vs particle log(Lorentz factor). The small symbols are from a GEANT4 simulation of the detector
configuration, and the large symbols are data from a direct accelerator measurement, as discussed in \cite{2003ICRC....4.2233S}.  Also
shown schematically is the relationship between the TRD signal resolution and the log(Lorentz factor).}\label{fig:TRDresp}
\end{figure}

The right panel of Figure \ref{fig:sigdist} shows the distribution of TRD signals for oxygen nuclei which are selected to be close to
minimum ionizing (using the Cherenkov detector).  These events are also selected to have more than six TRD tubes with a usable signal.
It is the width of this distribution (which remains essentially constant throughout the relativistic-rise region) that ultimately
determines the energy resolution of the TRD. The RMS ($1 \sigma$) width of the signal, S, distribution is $\Delta S\sim 8$ in these
units. As demonstrated in Figure \ref{fig:TRDresp}, which shows the response function of the TRD, this corresponds to a resolution in
log$_{10}(\gamma)\sim 0.3$, where $\gamma$ is the particle Lorentz Factor. This figure also shows the good agreement of the
GEANT4\cite{2003NIMPA.506..250G} detector simulation (small points) with the results of a test beam calibration of the same
configuration (large points).  This calibration is discussed in more detail elsewhere \cite{2003ICRC....4.2233S,2007NIMPA.579.1034A},
and has been shown to be consistent with the energies derived from the calorimeter signals in the $\sim$100 GeV/n region for carbon
and oxygen\cite{Maestro07}.

In practice, a simple least-squares energy fitter is used to determine the particle energy which best corresponds to
the signals measured in the TCD, CD and TRD.  For each event, a goodness-of-fit, $\chi_E^2$, is generated, based on the
known response functions of these detectors.

\subsection{Selection Cuts}
The data for this study were selected from time periods when the experiment was relatively stable and when there was no recent gas
venting, filling or other operations, and the atmospheric overburden was $\lesssim$5g/cm$^2$. Periods were also avoided when the high
voltage was being adjusted on any experimental system. These selections resulted in an effective data collection time of $\sim$21
days.

The events used in the final analysis must pass various identification and quality cuts.  The nuclear charge is selected by placing
cuts on the TCD and SCD signal distributions, similar to those shown in Figure \ref{fig:Scatters}.  An additional tracking cut helps
here by requiring that the calculated particle trajectory pass through the SCD within 2~cm of the pixel with the largest signal.  This
cut keeps only ``clean" events and eliminates possible albedo effects from the calorimeter just below the SCD.  Approximately 70\% of
Hi-Z triggers with TRD data survive this cut. To provide adequate signal resolution, only events with more than six analog samples in
the tube array are selected for analysis. In addition, the total track pathlength in the tubes is required to be at least 12~cm. These
two cuts admit $\sim$25\% of the events with a good level-two track fit. A consistency check between the $xz$-plane TRD signal and the
$yz$-view TRD signal (amounting to a simple cut on the distribution shown in Figure \ref{fig:sigdist}) further improves the overall
resolution, while accepting $\sim$65\% of the remaining events. A final cut is made on the likelihood of the track fit, rejecting poor
geometry events, while keeping over 90\% of those events which make it to this stage. The cumulative efficiency of all the cuts
applied to the events which trigger the instrument in this analysis is $\sim$10\%.

As mentioned above, the energy fitter produces a simple $\chi_E^2$ value which can be used to reject pathological events.  A loose
selection cut is made that keeps only those events with reasonable values of $\chi_E^2<10$.  This works rather well except when
knock-on electrons, or other processes, produce non-statistical upwards fluctuations in the CD signals. This effect can cause a
nucleus at low energy to mimic a high-energy event in the TRD.  This is easily understood by referring to the TRD response curve shown
in Figure \ref{fig:TRDresp}.  There is clear ambiguity in energy for a TRD signal of, for example, 115 scale units. The CD is needed
to determine on which side of minimum ionization the event lies. A cut is therefore placed on the CD to ensure the events are on the
high-energy side.

\begin{figure}
  \centering
  \includegraphics[width=0.99\linewidth]{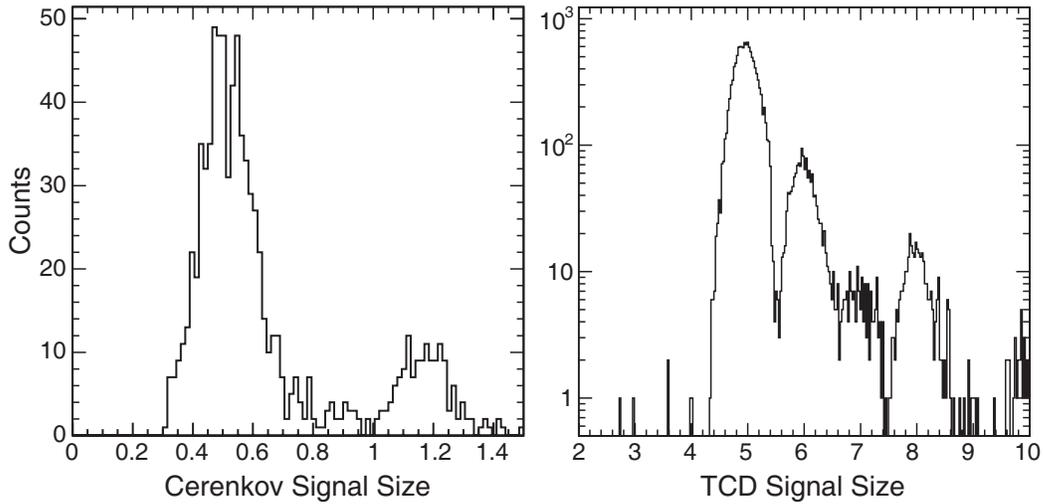}
  \caption{Left:~The distribution of Cerenkov signals from oxygen events selected to be between 115 and 125 on
   the response curve of Figure \ref{fig:TRDresp}.  See the text for details.  The right-hand population of events
    comprises particles near 100~GeV/n and the left-hand peak is a tail of
    low-energy events near 400~MeV/n which have passed identical cuts.  A cut
    at 1.05 provides effective discrimination between the two populations.
    Right:~The distribution of charge signals measured by the TCD for events
    which have an SCD charge of $Z=5.0\pm0.2$.  The effects of spallation are obvious as several of
    the nuclei enter the instrument with a charge larger than that measured in the SCD.}\label{fig:Cercut}
\end{figure}

The flight data themselves can be used to estimate the effectiveness of this Cherenkov cut. In Figure \ref{fig:Cercut} the CD signal
is plotted for oxygen nuclei which are selected as discussed above except for the Cherenkov cut illustrated in Figure
\ref{fig:Scatters}. These also are required to have a TRD signal between 30 and 35\% above minimum ionization.  According to Figure
\ref{fig:TRDresp}, these could either have an energy $\sim$400~MeV/n or $\sim$100~GeV/n from the response curve. It can be readily
seen that there are two easily separable populations - corresponding to the low-energy and high-energy events.  A simple cut on the CD
signal (above 1.05 in this case) ensures only the high-energy particle survive.

\subsection{Charge Overlap Rejection}

Reliably discriminating between different charges is extremely important for any experiment measuring secondary nuclei. For example,
because boron at high energy has only a few percent of the abundance of carbon, the peaks of the reconstructed populations of these
nuclei must be at least three standard deviations apart to prevent a significant background from carbon from leaking into the boron
sample.

With a TRD instrument, this requirement becomes even more stringent because the TRD signal response is not independent of Z (it scales
as Z$^2$). As a result, if a low-energy carbon nucleus gets misidentified by the charge detectors as a boron, it will not be counted
as a low-energy boron, but rather as a high-energy boron because of the incorrect Z$^2$ normalization factor.  This exacerbates the
``charge overlap background" problem because the steeply falling cosmic-ray spectrum means there are thousands more low-energy carbon
nuclei than high-energy boron nuclei.

Due to this additional effect, a separation of at least \textit{four} standard deviations between the boron and carbon populations is
required to make a background-free determination of the boron abundance. A similar condition applies to the measurements of nitrogen
and oxygen.

The CREAM instrument fortunately has two independent charge-determining detectors, the TCD and SCD, each of which has a resolution
significantly less than 0.2 charge units (over the range of charges considered here).  To combat the charge overlap problem, we place
circular cuts of radius $\Delta$Z=0.2 on all the charge peaks of the right panel of Figure \ref{fig:Scatters}.  This selection
produces effective cuts at $\Delta$Z=$\pm$0.14 on the summed charge distribution and completely removes interacting events.  The
resulting separation between the cut charge populations exceeds 4.5 standard deviations, thereby providing a clean sample of the
low-abundance boron and nitrogen nuclei at high energies.

\subsection{Energy Deconvolution}
All the events passing the above cuts are collected into energy bins of width log$_{10}$(E)=0.4. This corresponds to a width of $\sim
1.3 \sigma$ in terms of the energy resolution, discussed above.  Approximately 45,000 relativistic carbon nuclei and 40,000 oxygen
nuclei make it to this stage.

The true energy spectrum is derived from these binned values by correcting for energy spill-over between the bins due to the finite
instrumental resolution. To account for these effects in a system where the resolution is a function of energy, a simulation is
generally required.  For this measurement, the primary energy determination device is the TRD, so we use an accurate simulation of the
TRD and its fluctuations. This simulation uses a GEANT4 model of the upper CREAM payload including the TCD, TRD and SCD.  The TRD
component of this simulation has been discussed previously in \cite{2003ICRC....4.2233S}.

The events from this simulation are passed through the same analysis code that is used to evaluate and fit the flight data. Various
cross-checks are made to ensure the simulation provides an accurate representation of the data produced in the flight, including both
mean signal levels and the fluctuations of those signals.  Using the simulated results, a transfer matrix is obtained which can be
used to deconvolve the measured energy bins to produce estimated true energy bin populations.  This method, which has been discussed
in previous studies (\eg, \cite{1990ApJ...349..625S}), is an iterative process since, for accurate results, the spectral index used in
the simulation must match the (unknown) index of the data.  However, as noted by several previous authors
\cite{1974ApJ...191..331J,1994ApJ...429..736B}, the convergence of the iteration is rapid and small residual discrepancies have little
impact on the final result.

\section{Results and Corrections}
\begin{figure}[t]
  \centering
  \includegraphics[width=0.99\linewidth]{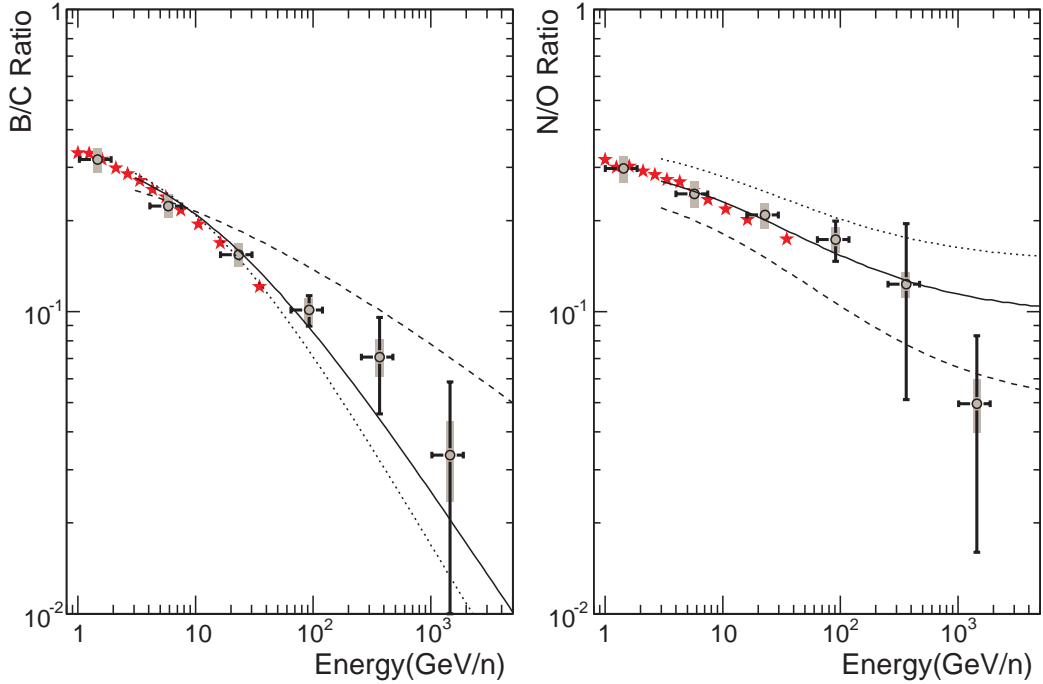}
  \caption{Measurements of the ratios of nuclei as a function of energy.
  Left:~Filled circles show the ratio of boron to carbon vs. energy after corrections.
  The horizontal errors are an estimate of the systematic error in the overall energy scale. The thin vertical lines
  correspond to the statistical error of the ratio and the grey bars show the systematic uncertainty in the ratio.  See text for details.
  The lines represent model calculations for various values of the magnetic-rigidity dependence
   parameter, $\delta$, in escape from the Galaxy - as discussed in the text. These are; solid line $\delta$=0.6, long-dashed line
    $\delta$=0.333, short-dashed line $\delta$=0.7. The stars are data from the space experiment, HEAO-3-C2\cite{1990A&A...233...96E}.
    Right: Filled circles show the ratio of nitrogen to oxygen vs. energy after corrections.  The error bars and data points are
    as in the left panel. The lines
     represent model calculations of this ratio with the escape parameter $\delta$=0.6 (solid line in the
     top left-hand panel). The different curves correspond to different assumptions on the amount of nitrogen in the source material.
     These are; solid line source N/O = 10\%, long-dashed line source N/O = 5\%, short-dashed line N/O = 15\%.}\label{fig:ratios}
\end{figure}

\begin{figure}[t]
  \centering
  \includegraphics[width=0.99\linewidth]{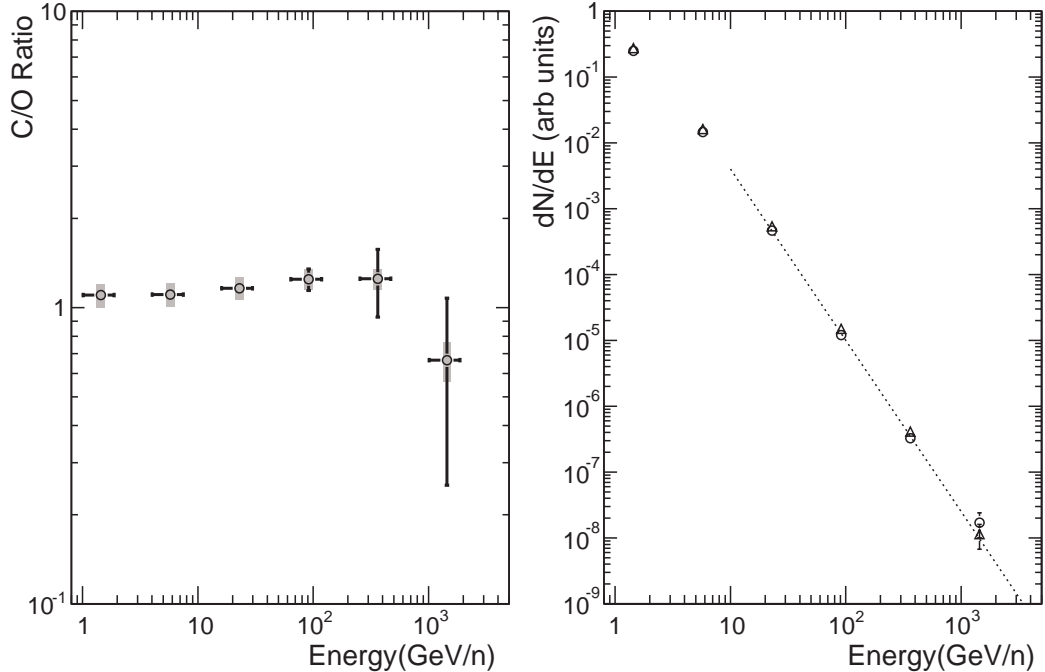}
  \caption{Left: Filled circles show the ratio of carbon to oxygen vs energy after corrections. Right: The
fluxes (energy/n$\times$area$\times$time$\times$solid angle)$^{-1}$ (arb. units) of carbon nuclei (open triangles) and oxygen nuclei
(open circles) from this measurement. The data are consistent with the dashed line, a power law $\propto$(Energy/n)$^{-\alpha}$ with a
spectral index $\alpha$=2.6. }\label{fig:coresults}
\end{figure}

Using the analysis, selection cuts, and deconvolution scheme outlined in the previous sections, the flight data are processed to
determine the ratios of elemental abundances as a function of energy into the TeV/n region. As described above, this is achieved with
a simple binning and deconvolution process which assigns the median energy of each bin by assuming a spectral index close to that of
each denominator element.  These results are given in Tables 1-3, and the associated Figures \ref{fig:ratios} and \ref{fig:coresults}.
Also shown for comparison are the results from the space measurement HEAO-3-C2\cite{1990A&A...233...96E} which used an assembly of
Cherenkov counters of different thresholds to measure elemental composition up to $\sim 35$GeV/n. These are the most statistically
significant measurements to date in the lower energy region. These tables show the abundance ratios B/C, N/O, and C/O, respectively,
as measured at the instrument as well as those values after extrapolation to the top of the atmosphere.

The corrections to the top of the atmosphere are made by considering separately the probability of charge-changing interactions in the
instrument and the atmosphere.  The right panel of Figure \ref{fig:Cercut} clearly shows that events that have fragmented can be
eliminated with charge cuts.  However, the efficiency of these cuts must be accounted for in the analysis procedure.

Charge-changing interactions also occur in the atmosphere above the instrument.  Obviously, these events cannot be identified, so a
post-analysis correction must be applied using nuclear interaction models.  For this purpose, we utilize both the partial
cross-section model of Tsao et al. \cite{1998ApJ...501..920T} and GEANT4 nuclear interaction simulations.  These corrections become
extremely important as the elemental abundance ratios decline.  For example, if 3\% of carbon nuclei (Z=6) change to boron (Z=5) in
the atmosphere above our instrument this is a fairly negligible correction at 5~GeV/n, where B/C$\sim$30\%.  At around $\sim1$TeV/n,
however, it becomes significant, since in this region B/C$\sim$5\%.

Since it appears that the B/C ratio continues to decline relatively steeply with energy, the accurate determination of
this ratio by balloon payloads will be intrinsically limited to the region below a few hundred GeV/n because of
uncertainties in these corrections. Measurements well above these energies can probably only be made successfully on a
future space mission, where there are no systematic limitations from atmospheric corrections.

There are several different uncertainties documented in the tables.  For the energy values, the quoted errors are systematic - arising
from the precision with which the data and simulation response curves can be aligned.  We estimate the fractional error in this
process to be $\sim$3\%.  In the response region where most of these data were collected, this precision corresponds to an
energy-scale uncertainty of $\sim$15\%, as displayed in the tables.

The entries for the corrected abundance ratios feature both statistical and systematic errors.  The statistical portion is derived in
the standard way by propagating the statistical errors of each element in a ratio, using the original counts of events in the bins
before deconvolution and before the atmospheric corrections.  The systematic errors in each ratio stem from two effects. The first
stems from residual uncertainty in the charge-dependent efficiency of the selection cuts used to construct the ratios.  There are
several cuts for which the charge dependence may amount to a few percent, so we use a conservative value of 10\% for the total
charge-dependent efficiency error. A second systematic effect derives from uncertainties in the atmospheric secondary corrections. The
partial cross-section values in the literature for carbon+Air $\rightarrow$ boron in the relativistic region can differ by up to
$\sim$30\%.   The typical correction in the ratio of B/C from atmospheric secondaries is of the order $\sim3$\%, and we assign a fixed
1\% systematic error for the uncertainty in correcting the ratios to the top of the atmosphere.  The systematic errors quoted in the
tables reflect whichever contribution dominates in the appropriate energy region.

To provide a check on these measurements we also present data on the C/O ratio in Figure \ref{fig:coresults}.  As both C and O are
primary nuclei - originating mostly in the cosmic-ray source - the expectation is that their abundance ratio should be close to
constant with energy. This expectation is in agreement with our measurements.

A further check can be seen in the right panel in Figure \ref{fig:coresults}.  Here the measured intensities of carbon and oxygen
nuclei are shown as a function of energy (these are differential elemental intensities given in arbitrary units since the overall
live-time of the flight has not yet been accurately determined). Apart from demonstrating that the CREAM measurements reported here
cover nearly 9 decades in differential intensity, these data show that the measured energy spectra for primary nuclei are consistent
with a differential energy spectral index of $\alpha$=2.6 at energies above 10GeV/n. This value is near the average of
previously-reported measurements and provides confidence that our analysis techniques and energy scales are accurate.

\section{Discussion}

The left panel of Figure \ref{fig:ratios} shows that the measured ratio of B/C declines fairly steeply with energy into the TeV/n
region.  Overlaid on this figure are three lines which represent the results of a simple leaky-box model of cosmic-ray propagation.
For each of the lines, a different value of $\delta$ is used.  The upper-most (long-dashed) line uses $\delta$=0.33, the solid line
uses $\delta$=0.6, and the short-dashed line uses $\delta$=0.7.  The small data points (stars) on the figure represent the ratio
measurements of the HEAO-C2-3 experiment\cite{1990A&A...233...96E}. The present data are consistent with these measurements and also
seem to be consistent with the $\delta$=0.6 curve.

Measurements of N/O allow one to investigate the apparent source abundance of nitrogen in cosmic rays, a topic which has been
discussed for many years. The cosmic nitrogen abundance is somewhat unusual in that it contains significant contributions from both
primary and secondary sources.  At a high-enough energy the primary component should come to dominate and the ratio N/O should flatten
out.  Our N/O measurements, again together with those of HEAO-C2-3 are shown in the right panel of Figure \ref{fig:ratios}. As before,
three simple leaky-box propagation models are overlaid on the data, in this case with the parameter $\delta$ fixed at 0.6.  The
different lines on this plot correspond to different assumptions on the source abundance of nitrogen compared to oxygen.  The solid
line corresponds to a source N/O = 10\%, the long-dashed line corresponds to a source N/O=5\% and the short-dashed line is for source
N/O = 15\%. Although lower-energy measurements on nitrogen isotopes have seemed to favor a source abundance N/O $\sim$5\%
(\eg,\cite{1985ApJS...57..173M}), these new data, which agree well with HEAO\cite{1990A&A...233...96E}, seem to prefer a source value
$\sim$10\%.

\section{Conclusions}
We have presented new measurements of the relative abundances of cosmic-ray secondary nuclei at high energies, where the energy
dependence of cosmic-ray propagation is relatively unexplored experimentally. These results have been produced with the first flight
of the CREAM high-altitude balloon instrument in Antarctica during austral summer 2004/2005. These are challenging measurements, and
the results span over three decades in energy and over nine decades in particle intensity. The energy scale for these results is
provided by a transition-radiation detector, operating mostly in the regime of the relativistic rise.  A critical part of the success
of this effort is the excellent resolution of the CREAM charge detectors. Both the TCD (plastic scintillator) and SCD (silicon
detector) have demonstrated charge resolutions of $\Delta$Z$<$0.2 charge units.  This provides excellent charge isolation between
primary nuclei and the much less abundant secondary nuclei at all energies.  Indeed, without this level of charge separation, the
background of misidentified nuclei would make this measurement difficult if not impossible above $\sim$100~GeV/n.

The results presented here show the interstellar propagation pathlength decreases fairly rapidly with energy, with an energy
dependence in the range $\delta\sim$0.5-0.6.  As a result, the propagation pathlength of cosmic nuclei is smaller by an order of
magnitude for particles in the TeV/n region compared to those at lower energies below 10~GeV/n. This high-energy pathlength ($\sim$1
g/cm$^2$), while small, is still large compared to the typical grammage of the Galactic disk ($\lesssim 0.002$ g/cm$^2$), and so it
does not significantly constrain residual pathlength models proposed for higher energies\cite{Swordy95-PL} or in the source
region\cite{2003A&A...410..189B}. The observations of N/O abundance ratios are consistent with this pathlength dependence and they
suggest an N/O source abundance close to 10\% - larger than some previous estimates based on lower-energy isotope measurements (see
e.g. \cite{1981ApJ...251L..27M}).

For this measurement, the dominant source of systematic error at high energy are uncertainties in the cross sections for producing
secondaries by charge-changing interactions in the atmosphere above the instrument. Because there is a significant decrease in the
interstellar pathlength with energy, the amount of boron, for example, at high energy is small - making the impact of uncertainties in
the atmospheric boron contribution significant above $\sim$100~GeV/n. At these energies the contribution from charge-changing
interactions in the atmosphere is similar in size to the total production of boron during propagation through the Galaxy. As a
consequence it seems likely that accurate measurements of B/C on high-altitude balloon experiments will be ultimately limited by
systematic errors in the TeV/n region until knowledge of these atmospheric corrections can be improved.

\section{Acknowledgements}
The work reported in this paper was funded by NASA research grants to the University of Maryland, the University of Chicago, Penn
State University, and the Ohio State University, by the Korean Ministry of Science and Technology and the Creative Research
Initiatives (MEMS Space Telescope) of MOST/KOSEF in Korea, and by INFN in Italy. The authors wish to acknowledge NASA/WFF for
provision and operation of flight support systems; Art Ruitberg, Suong Le, and Curtis Dunsmore of NASA/GSFC, and Carlos Urdiales of
Southwest Research Institute for assistance with HV design and potting; CERN for provision of excellent accelerator beams; the Fermi
National Accelerator Lab Thin Films Group for high-quality polishing and aluminization of optical elements; and Columbia Scientific
Ballooning Facility, National Science Foundation's Office of Polar Programs, and Raytheon Polar Services Company for outstanding
support of launch, flight and recovery operations in Antarctica. The TCD group acknowledges engineering contributions made by L.
Engel, J. Passaneau and S. Posey. The TRD group acknowledges engineering contributions made by R. Northrop, C. Smith and G.A.
Kelderhouse. The SCD group acknowledges contributions made by W. Han, H.J. Hyun, H.J. Kim, M.Y. Kim, K.W. Min, H. Park and K.I. Seon.

\newpage

\begin{center}
\begin{table}
\caption{Table of measurements of Boron to Carbon Ratio} \label{tab:results1}
\begin{tabular}
{|c|c|c|c|c|c|} \hline
Kinetic Energy& Ratio B/C & Ratio B/C & Energy Range\\
(GeV/n)& measured & corrected & (GeV/n)\\
\hline
1.4$\pm 0.2(sys.)$ &  0.353 & 0.320$\pm 0.007(stat.)\pm 0.03(sys.)$& 1-4\\
5.7$\pm 0.9(sys.)$ &  0.256 & 0.225$\pm 0.004(stat.)\pm 0.02(sys.)$ & 4-16\\
23$\pm 3(sys.)$ &  0.185  & 0.155$\pm 0.005(stat.)\pm 0.014(sys.)$ & 16-63\\
91$\pm 14(sys.)$ &  0.131  & 0.101$\pm 0.011(stat.)\pm 0.01(sys.)$ & 63-251\\
363$\pm 54(sys.)$ & 0.099  & 0.071$\pm 0.025(stat.)\pm 0.01(sys.)$ & 251-1000\\
1450$\pm 217(sys.)$ & 0.061 & 0.033$\pm 0.025(stat.)\pm 0.01(sys.)$ & 1000-4000\\
 \hline
\end{tabular}
\end{table}
\end{center}

\begin{center}
\begin{table}
\caption{Table of measurements of Nitrogen to Oxygen Ratio} \label{tab:results2}
\begin{tabular}
{|c|c|c|c|c|c|} \hline
Kinetic Energy& Ratio N/O & Ratio N/O & Energy Range\\
(GeV/n)& measured & corrected & (GeV/n)\\
\hline
1.4$\pm 0.2(sys.)$ &  0.341 & 0.299$\pm 0.006(stat.)\pm 0.03(sys.)$& 1-4\\
5.7$\pm 0.9(sys.)$ &  0.286 & 0.246$\pm 0.004(stat.)\pm 0.025(sys.)$ & 4-16\\
23$\pm 3(sys.)$ &  0.248  & 0.210$\pm 0.009(stat.)\pm 0.02(sys.)$ & 16-63\\
91$\pm 14(sys.)$ &  0.211  & 0.174$\pm 0.026(stat.)\pm 0.02(sys.)$ & 63-251\\
363$\pm 54(sys.)$ & 0.160  & 0.124$\pm 0.072(stat.)\pm 0.01(sys.)$ & 251-1000\\
1450$\pm 217(sys.)$ & 0.083 & 0.050$\pm 0.034(stat.)\pm 0.01(sys.)$ & 1000-4000\\
\hline
\end{tabular}
\end{table}
\end{center}

\begin{center}
\begin{table}
\caption{Table of measurements of Carbon to Oxygen Ratio} \label{tab:results3}
\begin{tabular}
{|c|c|c|c|c|c|} \hline
Kinetic Energy& Ratio C/O & Ratio C/O & Energy Range\\
(GeV/n)& measured & corrected & (GeV/n)\\
\hline
1.4$\pm 0.2(sys.)$ &  1.19 & 1.10$\pm 0.01(stat.)\pm 0.1(sys.)$& 1-4\\
5.7$\pm 0.9(sys.)$ &  1.19 & 1.11$\pm 0.01(stat.)\pm 0.1(sys.)$ & 4-16\\
23$\pm 3(sys.)$ &  1.25  & 1.16$\pm 0.03(stat.)\pm 0.1(sys.)$ & 16-63\\
91$\pm 14(sys.)$ &  1.34  & 1.25$\pm 0.10(stat.)\pm 0.1(sys.)$ & 63-251\\
363$\pm 54(sys.)$ & 1.35  & 1.25$\pm 0.32(stat.)\pm 0.1(sys.)$ & 251-1000\\
1450$\pm 217(sys.)$ & 0.71 & 0.66$\pm 0.41(stat.)\pm 0.1(sys.)$ & 1000-4000\\
\hline
\end{tabular}
\end{table}
\end{center}

\bibliography{CREAM_BIB}

\begin{thebibliography}{10}
\expandafter\ifx\csname url\endcsname\relax
  \def\url#1{\texttt{#1}}\fi
\expandafter\ifx\csname urlprefix\endcsname\relax\def\urlprefix{URL }\fi

\bibitem{2007ARNPS..57..285S}
A.~W. {Strong}, I.~V. {Moskalenko}, V.~S. {Ptuskin}, Annual Review of Nuclear
  and Particle Science 57 (2007) 285--327.

\bibitem{1972PhRvL..29..445J}
E.~{Juliusson}, P.~{Meyer}, D.~{M{\"u}ller}, \prl 29 (1972) 445--448.

\bibitem{1973ApJ...180..987S}
L.~H. {Smith}, A.~{Buffington}, D.~F. {Smoot}, L.~W. {Alvarez}, M.~A. {Wahlig},
  \apj 180 (1973) 987--1010.

\bibitem{1990ApJ...349..625S}
S.~P. {Swordy}, D.~{Mueller}, P.~{Meyer}, J.~{L'Heureux}, J.~M. {Grunsfeld},
  \apj 349 (1990) 625--633.

\bibitem{1990A&A...233...96E}
J.~J. {Engelmann}, P.~{Ferrando}, A.~{Soutoul}, P.~{Goret}, E.~{Juliusson},
  \aap 233 (1990) 96--111.

\bibitem{2000AIPC..528..421D}
A.~J. {Davis}, R.~A. {Mewaldt}, W.~R. {Binns}, E.~R. {Christian}, A.~C.
  {Cummings}, J.~S. {George}, P.~L. {Hink}, R.~A. {Leske}, T.~T. {von
  Rosenvinge}, M.~E. {Wiedenbeck}, N.~E. {Yanasak}, in: R.~A. {Mewaldt}, J.~R.
  {Jokipii}, M.~A. {Lee}, E.~{M{\"o}bius}, T.~H. {Zurbuchen} (Eds.),
  Acceleration and Transport of Energetic Particles Observed in the
  Heliosphere, Vol. 528 of American Institute of Physics Conference Series,
  2000, pp. 421--+.

\bibitem{2005ICRC-CREAM}
E.-S. {Seo}, et~al., in: Proc. 29th ICRC (Pune), Vol.~3, 2005, pp. 101--104.

\bibitem{2007NIMPA.579.1034A}
H.~S. {Ahn}, P.~{Allison}, M.~G. {Bagliesi}, J.~J. {Beatty}, G.~{Bigongiari},
  P.~{Boyle}, J.~T. {Childers}, N.~B. {Conklin}, S.~{Coutu}, M.~A. {Duvernois},
  O.~{Ganel}, J.~H. {Han}, J.~A. {Jeon}, K.~C. {Kim}, J.~K. {Lee}, M.~H. {Lee},
  L.~{Lutz}, P.~{Maestro}, A.~{Malinin}, P.~S. {Marrocchesi}, S.~A. {Minnick},
  S.~I. {Mognet}, S.~W. {Nam}, S.~L. {Nutter}, I.~H. {Park}, N.~H. {Park},
  E.~S. {Seo}, R.~{Sina}, S.~P. {Swordy}, S.~P. {Wakely}, J.~{Wu}, J.~{Yang},
  Y.~S. {Yoon}, R.~{Zei}, S.~Y. {Zinn}, Nuclear Instruments and Methods in
  Physics Research A 579 (2007) 1034--1053.

\bibitem{2006NuPhS.150..272A}
H.~S. {Ahn}, M.~G. {Bagliesi}, J.~J. {Beatty}, G.~{Bigongiari},
  A.~{Castellina}, J.~T. {Childers}, N.~B. {Conklin}, S.~{Coutu}, M.~A.
  {Duvernois}, O.~{Ganel}, J.~H. {Han}, H.~J. {Hyun}, T.~G. {Kang}, H.~J.
  {Kim}, K.~C. {Kim}, M.~Y. {Kim}, T.~{Kim}, Y.~J. {Kim}, J.~K. {Lee}, M.~H.
  {Lee}, L.~{Lutz}, P.~{Maestro}, A.~{Malinine}, P.~S. {Marrocchesi}, S.~I.
  {Mognet}, S.~W. {Nam}, S.~{Nutter}, N.~H. {Park}, H.~{Park}, I.~H. {Park},
  E.~S. {Seo}, R.~{Sina}, S.~{Syed}, C.~{Song}, S.~{Swordy}, J.~{Wu},
  J.~{Yang}, H.~Q. {Zhang}, R.~{Zei}, S.~Y. {Zinn}, Nuclear Physics B
  Proceedings Supplements 150 (2006) 272--275.

\bibitem{2007NIMPA.570..286P}
I.~H. {Park}, N.~H. {Park}, S.~W. {Nam}, et~al., Nuclear Instruments and
  Methods in Physics Research A 570 (2007) 286--291.

\bibitem{2002APh....18...67W}
S.~P. {Wakely}, APh 18 (2002) 67--87.

\bibitem{WalentaRelRise79}
A.~H. Walenta, J.~Fischer, H.~Okuno, C.~L. Wang, Nuclear Instruments and
  Methods 161~(1) (1979) 45--58.

\bibitem{2003ICRC....4.2233S}
S.~P. {Swordy}, P.~{Boyle}, S.~{Wakely}, in: Proc. 28th ICRC (Tskuba), 2003,
  pp. 2233--2236.

\bibitem{2007ICRC-TCD}
T.~J. {Brandt}, et~al., in: Proc. 30th ICRC (Merida), 2007.

\bibitem{2003NIMPA.506..250G}
{GEANT4 Collaboration}, S.~{Agostinelli}, J.~{Allison}, K.~{Amako},
  J.~{Apostolakis}, H.~{Araujo}, P.~{Arce}, M.~{Asai}, D.~{Axen},
  S.~{Banerjee}, G.~{Barrand}, F.~{Behner}, L.~{Bellagamba}, J.~{Boudreau},
  L.~{Broglia}, A.~{Brunengo}, H.~{Burkhardt}, S.~{Chauvie}, J.~{Chuma},
  R.~{Chytracek}, G.~{Cooperman}, G.~{Cosmo}, P.~{Degtyarenko},
  A.~{dell'Acqua}, G.~{Depaola}, D.~{Dietrich}, R.~{Enami}, A.~{Feliciello},
  C.~{Ferguson}, H.~{Fesefeldt}, G.~{Folger}, F.~{Foppiano}, A.~{Forti},
  S.~{Garelli}, S.~{Giani}, R.~{Giannitrapani}, D.~{Gibin}, J.~J. {G{\'o}mez
  Cadenas}, I.~{Gonz{\'a}lez}, G.~{Gracia Abril}, G.~{Greeniaus}, W.~{Greiner},
  V.~{Grichine}, A.~{Grossheim}, S.~{Guatelli}, P.~{Gumplinger}, R.~{Hamatsu},
  K.~{Hashimoto}, H.~{Hasui}, A.~{Heikkinen}, A.~{Howard}, V.~{Ivanchenko},
  A.~{Johnson}, F.~W. {Jones}, J.~{Kallenbach}, N.~{Kanaya}, M.~{Kawabata},
  Y.~{Kawabata}, M.~{Kawaguti}, S.~{Kelner}, P.~{Kent}, A.~{Kimura},
  T.~{Kodama}, R.~{Kokoulin}, M.~{Kossov}, H.~{Kurashige}, E.~{Lamanna},
  T.~{Lamp{\'e}n}, V.~{Lara}, V.~{Lefebure}, F.~{Lei}, M.~{Liendl},
  W.~{Lockman}, F.~{Longo}, S.~{Magni}, M.~{Maire}, E.~{Medernach},
  K.~{Minamimoto}, P.~{Mora de Freitas}, Y.~{Morita}, K.~{Murakami},
  M.~{Nagamatu}, R.~{Nartallo}, P.~{Nieminen}, T.~{Nishimura}, K.~{Ohtsubo},
  M.~{Okamura}, S.~{O'Neale}, Y.~{Oohata}, K.~{Paech}, J.~{Perl},
  A.~{Pfeiffer}, M.~G. {Pia}, F.~{Ranjard}, A.~{Rybin}, S.~{Sadilov}, E.~{di
  Salvo}, G.~{Santin}, T.~{Sasaki}, N.~{Savvas}, Y.~{Sawada}, S.~{Scherer},
  S.~{Sei}, V.~{Sirotenko}, D.~{Smith}, N.~{Starkov}, H.~{Stoecker},
  J.~{Sulkimo}, M.~{Takahata}, S.~{Tanaka}, E.~{Tcherniaev}, E.~{Safai
  Tehrani}, M.~{Tropeano}, P.~{Truscott}, H.~{Uno}, L.~{Urban}, P.~{Urban},
  M.~{Verderi}, A.~{Walkden}, W.~{Wander}, H.~{Weber}, J.~P. {Wellisch},
  T.~{Wenaus}, D.~C. {Williams}, D.~{Wright}, T.~{Yamada}, H.~{Yoshida},
  D.~{Zschiesche}, Nuclear Instruments and Methods in Physics Research A 506
  (2003) 250--303.

\bibitem{Maestro07}
P.~Maestro, et~al., in: Proc. 30th ICRC (Merida), 2007.

\bibitem{1974ApJ...191..331J}
E.~{Juliusson}, \apj 191 (1974) 331--348.

\bibitem{1994ApJ...429..736B}
J.~{Buckley}, J.~{Dwyer}, D.~{Mueller}, S.~{Swordy}, K.~K. {Tang}, \apj 429
  (1994) 736--747.

\bibitem{1998ApJ...501..920T}
C.~H. {Tsao}, R.~{Silberberg}, A.~F. {Barghouty}, \apj 501 (1998) 920--+.

\bibitem{1985ApJS...57..173M}
J.-P. {Meyer}, \apjs 57 (1985) 173--204.

\bibitem{Swordy95-PL}
S.~{Swordy}, in: Proc. 24th ICRC (Rome), 1995, p. 697.

\bibitem{2003A&A...410..189B}
E.~G. {Berezhko}, L.~T. {Ksenofontov}, V.~S. {Ptuskin}, V.~N. {Zirakashvili},
  H.~J. {V{\"o}lk}, \aap 410 (2003) 189--198.

\bibitem{1981ApJ...251L..27M}
R.~A. {Mewaldt}, J.~D. {Spalding}, E.~C. {Stone}, R.~E. {Vogt}, \apjl 251
  (1981) L27--L31.

\end{thebibliography}
\bibliographystyle{elsart-num}


\end{document}